\documentclass[aip,rsi,reprint,graphicx]{revtex4-1} 

\usepackage{graphicx}
\usepackage{gensymb}
\usepackage{txfonts}
\usepackage{adjustbox}
\usepackage{multirow}
\usepackage{epstopdf}
\usepackage{tabulary}
\usepackage{tabularx}
\usepackage{hhline}
\usepackage{booktabs}
\usepackage{array}
\usepackage{subcaption}
\usepackage[utf8]{inputenc}
\usepackage{float}
\begin{document}


\title{\textbf{NanoRocks: Design and Performance of an Experiment Studying Planet Formation on the International Space Station}} 

\author{Julie Brisset}
\email{julie.brisset@ucf.edu}
\author{Joshua Colwell}
\author{Adrienne Dove}
\author{Doug Maukonen}
\affiliation{Center for Microgravity Research, Department of Physics and Florida Space Institute, University of Central Florida, 4111 Libra Drive, Orlando FL-32816, USA, Phone: +1-407-823-6168}

\date{\today}

\begin{abstract}

In an effort to better understand the early stages of planet formation, we have developed a 1.5U payload that flew on the International Space Station (ISS) in the NanoRacks NanoLab facility between September 2014 and March 2016. This payload, named NanoRocks, ran a particle collision experiment under long-term microgravity conditions. The objectives of the experiment were (a) to observe collisions between mm-sized particles at relative velocities of  $<$1~cm/s, and (b) to study the formation and disruption of particle clusters for different particle types and collision velocities. Four types of particles were used: mm-sized acrylic, glass, and copper beads, and 0.75 mm-sized JSC-1 lunar regolith simulant grains. The particles were placed in sample cells carved out of an aluminum tray. This tray was attached to one side of the payload casing with three springs. Every 60~s, the tray was agitated and the resulting collisions between the particles in the sample cells were recorded by the experiment camera. 

During the 18 months the payload stayed on ISS, we obtained 158 videos, thus recording a great number of collisions. The average particle velocities in the sample cells after each shaking event were around 1 cm/s. After shaking stopped, the inter-particle collisions damped the particle kinetic energy in less than 20~s, reducing the average particle velocity to below 1 mm/s, and eventually slowing them to below our detection threshold. As the particle velocity decreased, we observed the transition from bouncing to sticking collisions. We recorded the formation of particle clusters at the end of each experiment run. This paper describes the design and performance of the NanoRocks ISS payload.

\end{abstract}


\maketitle 

\section{Introduction}
\label{s:intro}
\vspace{-5pt}
The current state of knowledge on the early stages of planet formation includes the growth of dust grains around newly formed stars inside of protoplanetary disks (PPDs) \citep{weidenschilling_cuzzi1993PP3,chambers2004}. These grains are observed to be about a micrometer in size after condensation from the gaseous phase \citep{dalessio2001, vanboekel2005}. Numerical simulations and laboratory experiments have shown that they then proceed to grow inside the PPD through sticking collisions \citep{weidenschilling_cuzzi1993PP3, testi_et_al2014PP}, reaching mm to cm sizes. The progress of grain growth can be observed until this stage in the latest ALMA (Attacama Large Millimeter Array) pictures \citep{macgregor2013}. However, growth through sticking cannot account for the formation of m- to km-sized bodies, the so-called planetesimals. Indeed, at sizes of a mm to a cm, the grain relative velocities in the PPD are expected to increase \citep{weidenschilling1977MNRAS}, and collisions between particles cease to result in sticking. As shown in previous laboratory experiments \citep{blum_wurm2008ARAA}, these particle sizes and velocities also lead to bouncing and fragmentation of the collision partners. Numerical simulations have shown that these collisional outcomes can stall the growth of particles inside the PPD, a phenomenon called the ``bouncing barrier'' \citep{zsom_et_al2010AA}. 

Several planetesimal formation processes have been put forward to reconcile this bouncing barrier with the observed presence of planetesimal remnants such as asteroids and comets in our Solar System. In particular, Johansen et al. (2014) \cite{johansen2014} proposed that gas turbulences and instabilities inside the PPD lead to concentrations of mm- to cm-sized particles into dense clouds, which eventually collapse under their own weight, forming a m- to km-sized planetesimal. The collisional behavior of dust particles at the bouncing barrier and inside multi-particle environments, such as these concentrated clouds, therefore plays an essential role in the processes leading to planetesimal formation.

In an effort to better understand the early stages of planet formation, dust collision experiments were performed around the sticking to bouncing transition. Weidling et al. (2011) \citep{weidling_et_al2012Icarus} studied bouncing collisions between mm-sized SiO$_2$ aggregates and Kothe et al. (2013) \citep{kothe_et_al2013Icarus} and Brisset et al. (2017) \citep{brisset2017} observed the collisional behavior of 100-$\mu$m dust aggregates. In order to achieve the range of relative velocities between colliding particles relevant to the bouncing transition (1~cm/s and below), all of these experiments were performed under microgravity conditions. The platform used to run these experiments was the drop tower in Bremen, Germany, which allows microgravity experiments of 9~s duration.

We designed the NanoRocks experiment to use the NanoRacks platform on the International Space Station (ISS) to perform a long-duration microgravity experiment. In its 1.5U format (10$\times$10$\times$15 cm$^3$) it recorded the collisional behavior of several types of particles during repeated 5-minute experiment runs over a period of  18~months between October 2014 and March 2016. In this paper, we describe the design and performance of the experiment. In Section~\ref{s:setup}, we describe the NanoRocks experimental setup. In Section~\ref{s:run} we discuss the performance of the experiment onboard the ISS, and our conclusion are in Section~\ref{s:conclusion}. Scientific results of the experiment will be published in a separate paper.
\vspace{-7pt} 
\enlargethispage{\baselineskip}

\begin{figure}[!htbp]
  \begin{center}
  \includegraphics[width = 0.48\textwidth]{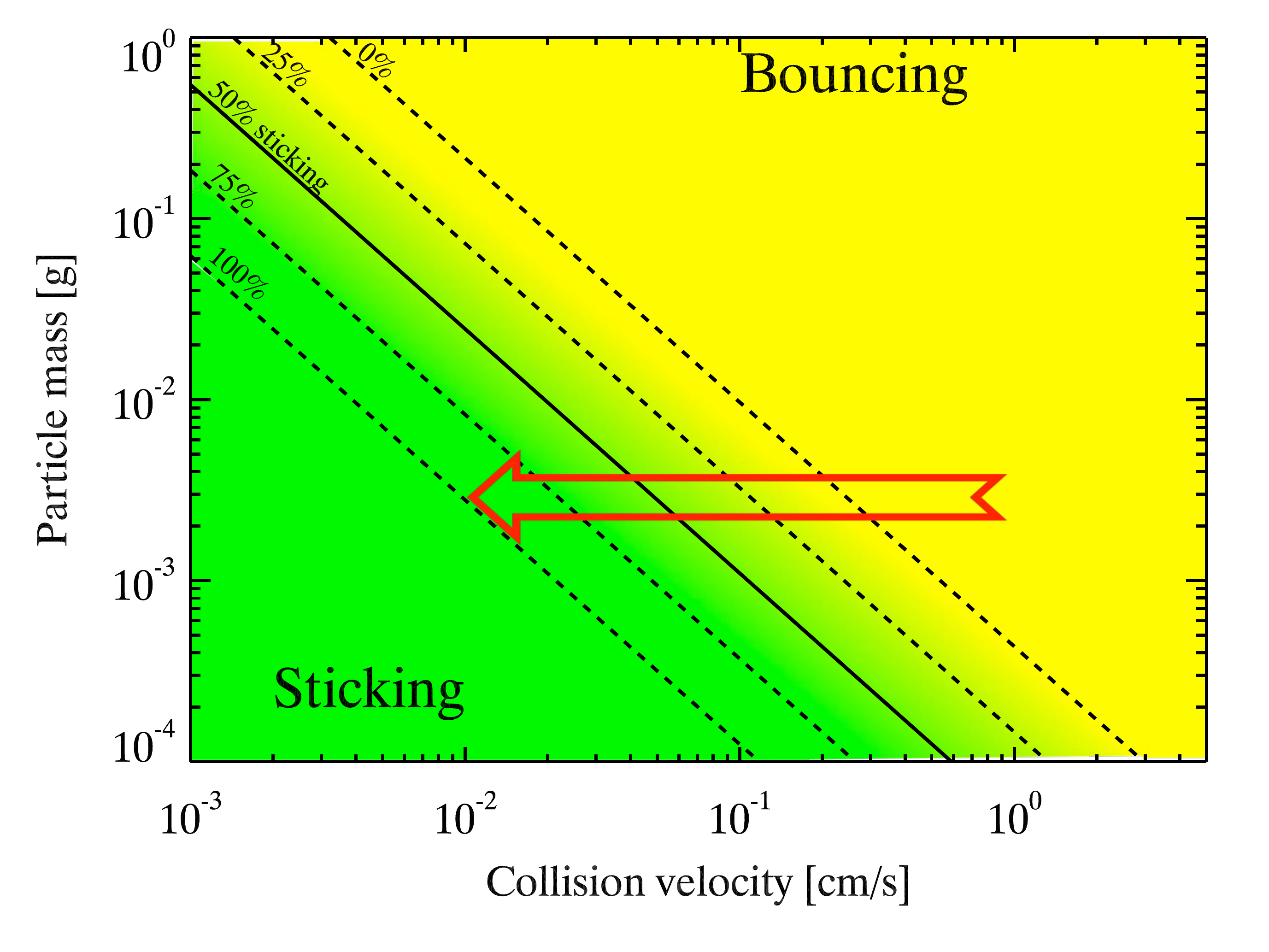}
 \caption{Dust collision model developed by G\"{u}ttler et al. (2009) and refined by Kothe, et al. (2013) \citep{guettler_et_al2009ApJ,kothe_et_al2013Icarus}. The outcome of collisions between two dust particles is depicted according to the particle mass and relative velocities. Green: collisions result in sticking, yellow: collisions result in bouncing. The transition from sticking to bouncing is marked by 5 sticking probablility lines (0, 25, 50, 75 and 100 \% sticking probablility). The parameter space studied with the NanoRocks experiment is marked by the entire length of the red arrow: the particles have masses of $\sim$ 5~mg, and their relative velocities decrease from $\sim$1~cm/s to $<$0.1~mm/s during each experiment run, thus transitioning from bouncing collisions to sticking collisions.}
 \label{f:model}
 \end{center}
\end{figure}

\section{The NanoRocks Experiment}
\label{s:setup}
\vspace{-7pt}

\subsection{Scientific Background}
\label{s:science}
\vspace{-7pt}

The objective of the NanoRocks experiment was to study low-energy collisions of mm-sized particles of different shapes and materials. In particular, the experiment was designed to study the bouncing-to-sticking transition for collisions with decreasing collision velocity. Current state-of-the-art research on the very early stages of planet formation relies on the understanding of dust particle behavior upon collisions inside the PPD. G\"{u}ttler et al. (2009) \citep{guettler_et_al2009ApJ} have developed a dust collision model predicting the outcome of collisions between dust aggregates that depends on their masses and relative velocities. The model predicts that small particles colliding at low relative velocities always stick to each other (``hit-and-stick'' behavior \cite{blum_et_al1998EM&P}). However, with increasing particle mass and relative velocities, collision outcomes transition to bouncing. Figure~\ref{f:model} shows the collision outcomes for dust aggregates of masses from 10$^{-4}$ to 1 grams and relative velocities from 10$^{-3}$ to 10 cm/s. This region of the parameter field covers the transition between sticking (green) and bouncing (yellow). \\
\indent The NanoRocks parameter space is indicated by the entire length of the red arrow in Figure~\ref{f:model}. The low relative velocities required for these collisions (from a few cm/s to under a mm/s) can only be obtained under long-term microgravity conditions. Residual accelerations up to 10$^{-4}g$, with $g$ being the acceleration of gravity at the surface of the Earth, were acceptable for the intended data collection. For these reasons, the ISS was the ideal platform to fly the NanoRocks experiment.

\begin{figure}[tp]
  \begin{center}
  \includegraphics[width = 0.48\textwidth]{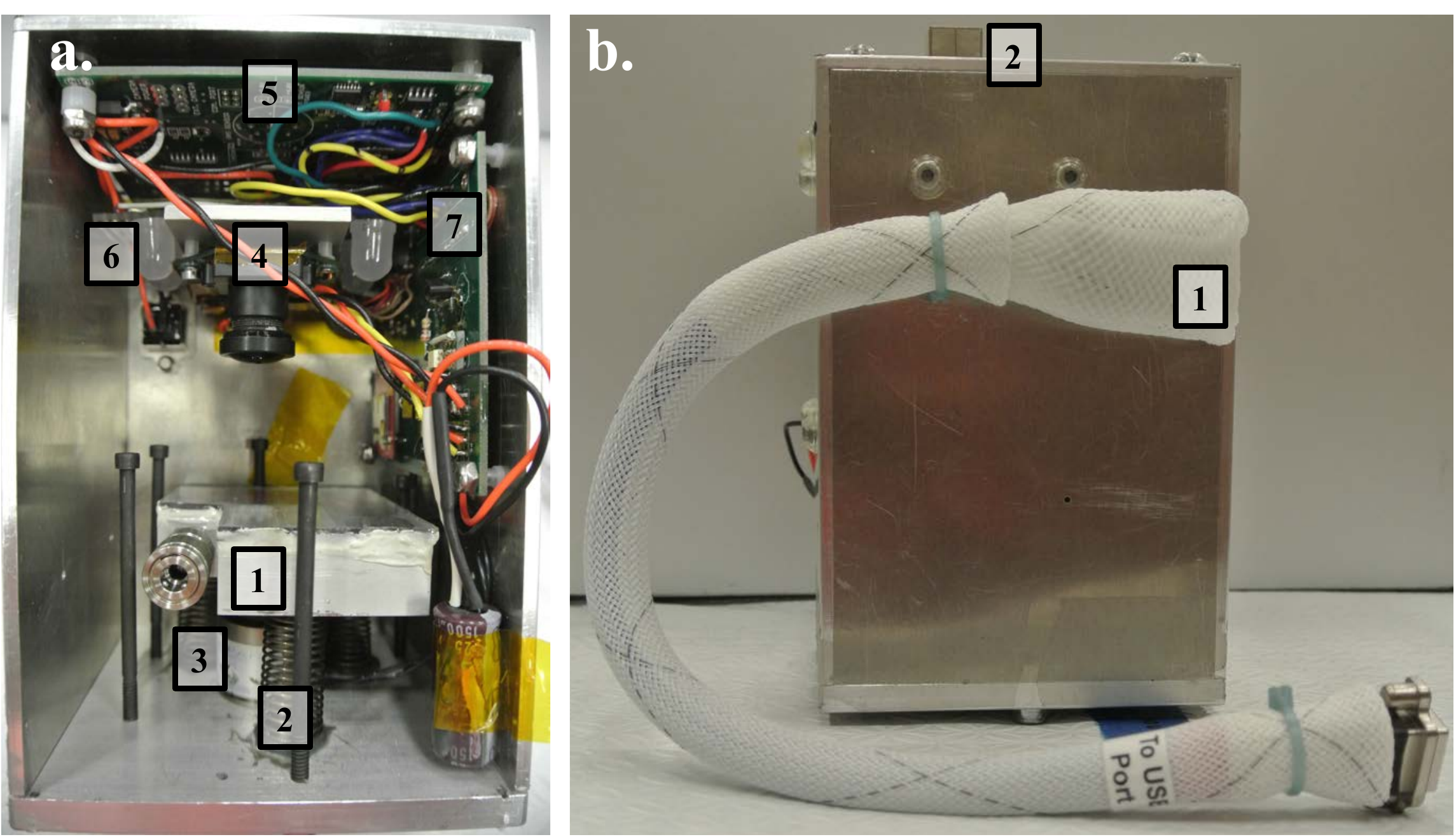}
 \caption{The NanoRocks flight unit: (a) Interior view showing the experiment tray (from the side) (1) attached to the bottom of the casing via springs (2). Under the tray, the shaking magnet can be seen (3). Above the tray, the HackHD camera (4) is placed under the NanoRacks Control Module (NCM) board (5) and surrounded by two LED arrays (6). The UCF custom electronics board is accommodated on the right side of the experiment casing (7). (b) Exterior view showing the USB data umbilical from the experiment to the additional power (1). The USB port on top of the experiment provided primary power and a connection the ISS laptop (2).}
 \label{f:flight_unit}
 \end{center}
\end{figure}

\subsection{Interfacing with the NanoRacks ISS Payload Platform}
\label{s:NanoRacks}
The NanoRacks NanoLab provides power (5 V at 400 mA per line) and data routing to ground. The NanoRocks payload was allocated a 1.5U and 1.5~kg mass limit. Both power and data were provided via USB 2.0 connections. Further details on the payload requirements to fly on the NanoRacks NanoLab can be found in the NanoRacks Internal Platforms 1A/2A and NanoLab Modules Interface Control Document \citep{NanoRacksICD}.

All hardware and flight vehicle compatibility tests were provided and performed by NanoRacks. Figure~\ref{f:flight_unit}b shows an exterior view of the NanoRocks experiment flight unit with its USB connections. For this experiment, the use of two connectors provided additional power for the camera. Once on the ISS, data exchange with the payload was possible via NanoRacks. Video data downloaded from the ISS laptop was made available via an ftp repository. Command file upload was also possible via NanoRacks.

\begin{figure*}[t]
  \begin{center}
  \includegraphics[width = 0.95\textwidth]{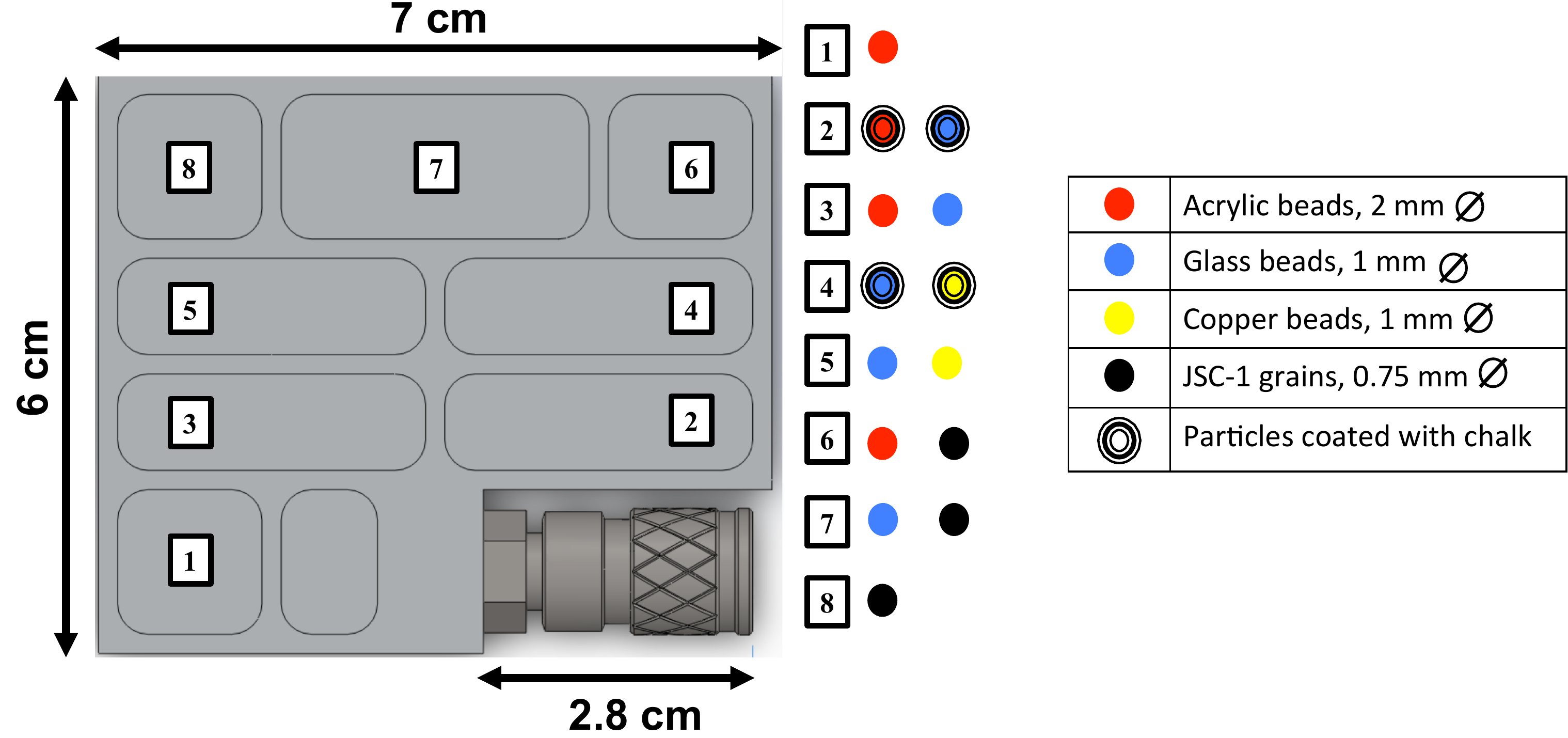}
 \caption{The 8 NanoRocks sample cells and their contents, according to the legend to the right. The tray is shown from the top (as seen by the experiment camera). At the bottom right is the vacuum valve that allowed for the tray evacuation before flight.}
 \label{f:particles}
 \end{center}
\end{figure*}


\subsection{Experiment Setup}

The main component of the experiment is an aluminum tray ($\sim$8$\times$8$\times$2~cm$^3$) which was divided into eight sample cells each holding different types and combinations of particles (Figure~\ref{f:particles}). Each sample cell was 3~mm deep, while the other dimensions varied (Figure~\ref{f:particles}). The tray was sealed with a transparent polycarbonate top plate to allow for observation of the particles inside the sample cells. The tray was evacuated via an attached vacuum valve to $\sim$10$^{-4}$ bar. The evacuation of the experiment tray was performed at the Center for Microgravity Research (CMR) before shipping the payload to NanoRacks for integration. During the development of the experiment, the particle tray was tested to ensure that it could be safely evacuated, and to determine if it would maintain vacuum during its operation on station.  While we were able to achieve high vacuum while the tray was connected to a pump ($<10^{-6}$ bar), the tray did not hold this quality of vacuum once disconnected, stabilizing around $10^{-4}$ bar. Our longest test revealed that, after several days, the pressure in the tray was on the order of $10^{-2}$ bar. Most of the increase in pressure came within the first day after it was removed from vacuum, and naturally slowed after that. Therefore, we expect that the tray gradually leaked, increasing the pressure over the course of the flight, but no measurements of the tray pressure were possible during the mission.

\begin{figure*}[t]
  \centering
  \begin{minipage}{0.5\textwidth}
    \includegraphics[width=1\textwidth]{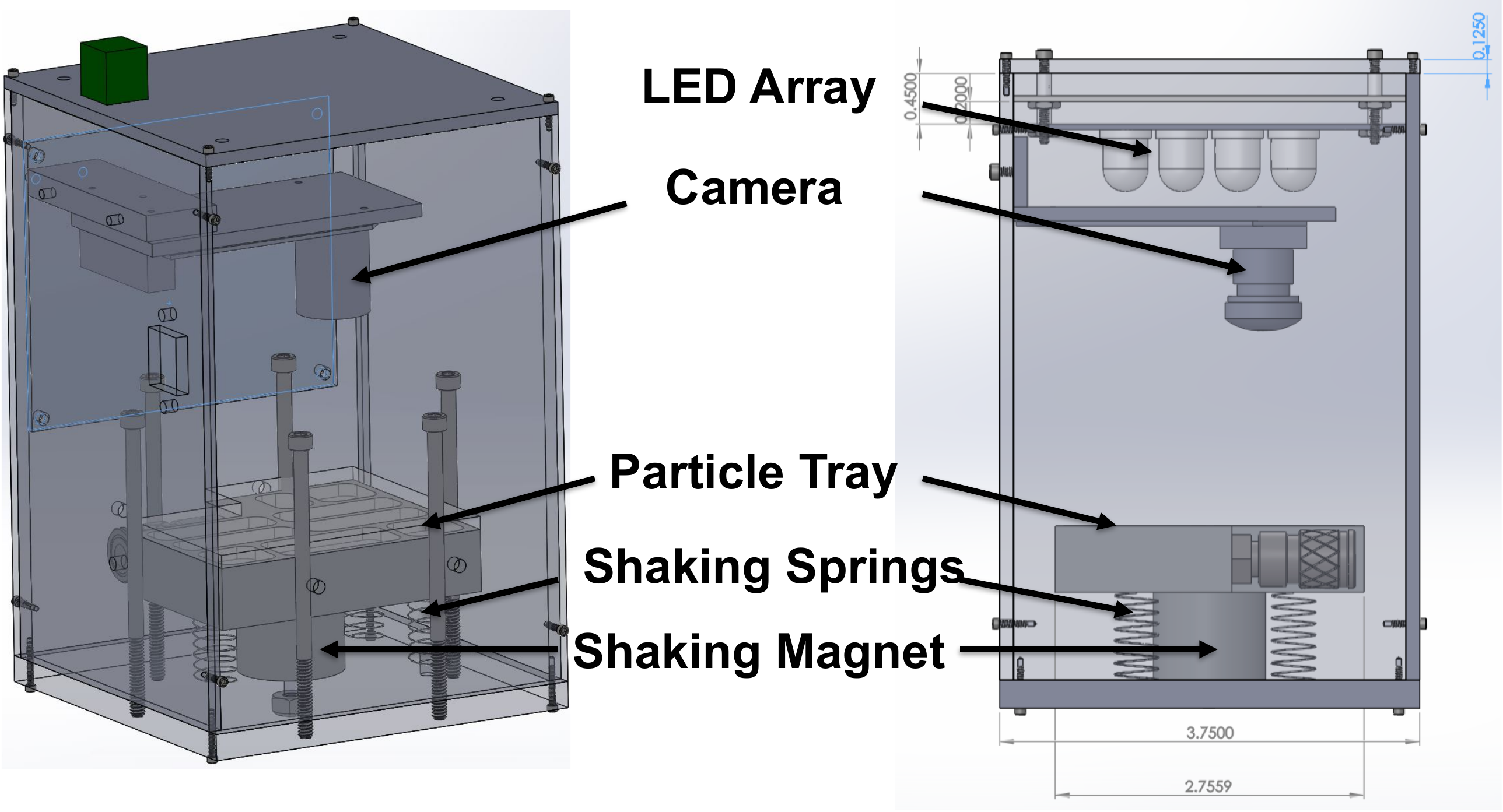}\
    \caption{The NanoRocks hardware design. The particle tray is the main component of the experiment and contains the particle samples. Above it, the LED arrays and camera allow for video recording during an experiment run. The springs and magnet below the tray created particle agitation.}
    \label{f:design}
  \end{minipage}%
  \hfill
  \begin{minipage}{0.48\textwidth}
    \includegraphics[width=0.69\textwidth]{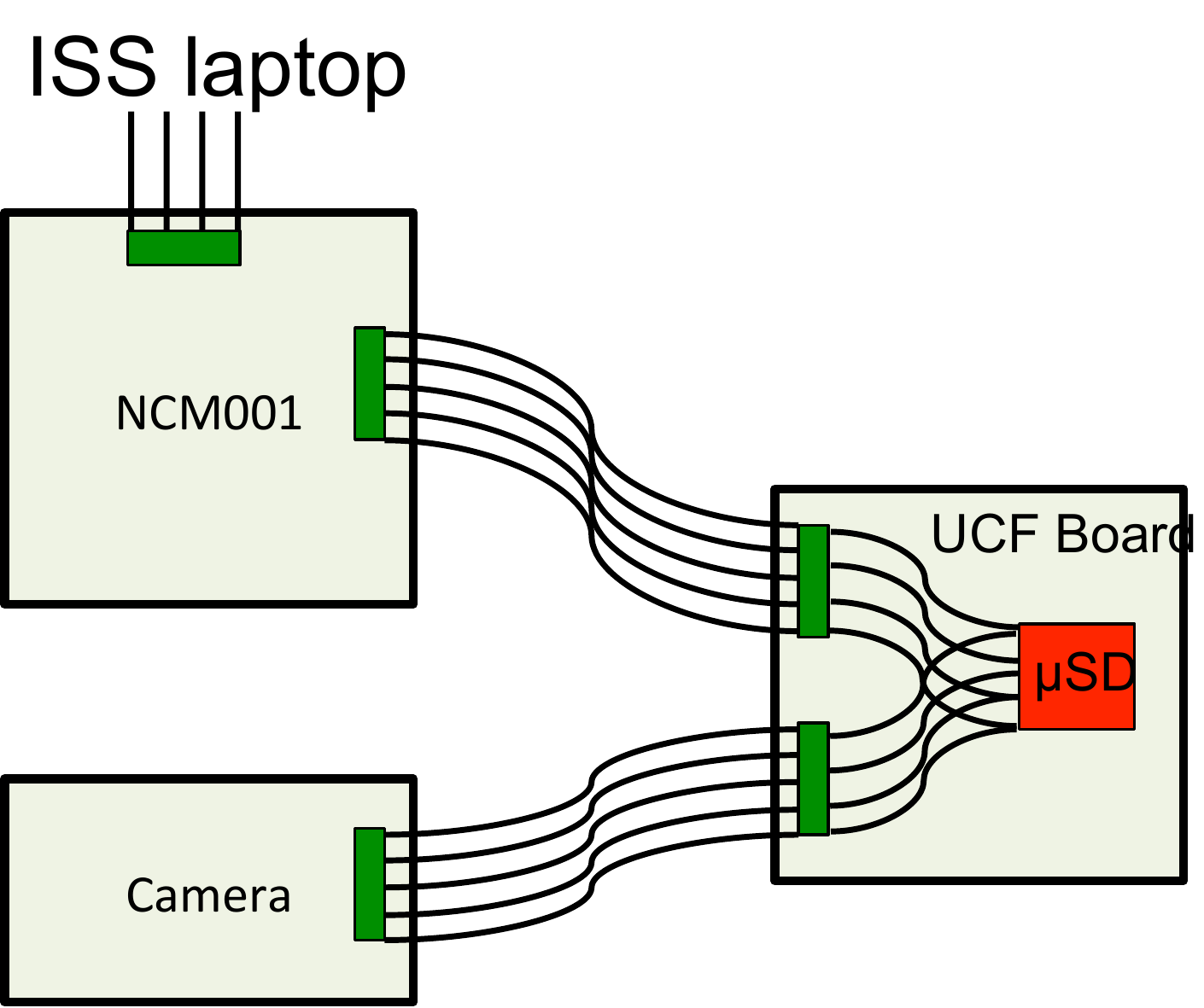}
     \caption{A schematic electronics diagram illustrating the switching role of the UCF custom electronics board. The experiment's micro-SD card was on the UCF board, whcih switched it alternatively to the camera for data collection and to the NCM board for data download to the ISS laptop.}
      \label{f:electronics}
  \end{minipage}
\end{figure*}

The experiment tray was mounted on three springs to allow for shaking to induce particle velocities via collisions with the walls. During an experiment run, an electromagnet grabbed and released the bottom of the tray every 60~s, resulting in shaking of the particle samples. An array of 4 LEDs (2 on each side of the camera) provided the required illumination, while a Hack HD camera \citep{HackHD} recorded the motion of the particles (Figure~\ref{f:design}). A light diffuser for the LED array was originally planned in the form of a layer of blurred paper between the LEDs and the experiment tray. Upon completion of hardware assembly this diffuser was not implemented as it could not be fixed reliably to the side walls and a displacement of the diffuser sheet during handling or launch would have significantly impaired scientific data collection. Losing partial data return through LED glare ($<$ 5 \% of the observed area) was considered an acceptable mitigation to the risk of losing the entire science data return. The video recording was performed at a resolution of 1080P and 30 fps, allowing for the observation of particle motion from one frame to the next and the determination of collision parameters. The experiment data recorded during an experiment run was stored on the experiment memory card, a 32GB microSDHC class 4 card. NanoRocks was also outfitted with its own electronics to allow for autonomous operation of the experiment during its stay on ISS.

\vspace{-15pt}

\subsection{Electronics and Data Collection}

The NanoRocks electronics were composed of two boards: the NanoRacks Control Module 001 (NCM) provided by Celestial Circuits and a custom electronics board produced at UCF. The NCM provided power switching to the different electronic components (the camera, electro-magnet, and LED array) as well as a data connection to the ISS laptop, which could read the microSD card on the board. For use of the NCM board with its compatible camera, no additional electronics would have been required. However, during the design and testing of the hardware at the CMR, we recognized that the resolution and frame rate provided by the NCM camera were not sufficient for the NanoRocks science goals. Therefore, an alternative camera, the HackHD 1080P, was chosen for the experiment. The HackHD can record frames autonomously to its own micro SD card upon power on, following a script loaded on the card. 
However, it was not possible to read the data collected by the camera on its SD card via the NCM board, as the SD card was inserted in the camera card reader which had no connection to the NCM card reader. Therefore it was necessary to add a custom board that served as a switch for the micro SD card. The experiment card was physically inserted into the UCF board but connected to both the camera and the NCM board via the appropriate data lines and read alternately by each of them. 

Before the start of an experiment run, the microSD card was connected to the camera for data recording. Once the experiment run was completed, the UCF board switched the card back to the NCM board, which was then able to transfer the recorded data to the ISS laptop. The experiment sequence (see Section~\ref{s:exp_sequence}) was run by the NCM board, reading the experiment parameters from a command text file. Every power cycle of the payload reinitiated the NCM, thus restarting an experiment sequence. It was possible to control the experiment parameters from one run to the next by uploading an updated parameter file to the NCM before a payload power cycle. The parameters included in the command file were the total number of shaking events and the time lapse between two shaking events, thus determining the length of the experiment run.

The recording parameters such as frame rate, resolution, and recording mode were set by a text file on the micro SD card. As it was not possible to update that file autonomously through the experiment electronics, it was prepared before payload delivery and thus determined the recording parameters for the entire flight. The resolution was set in order to distinguish particles and features down to 75~$\mu$m, which might result from the breakup of the smallest JSC-1 aggregates during collisions. With an experiment tray side of 8~cm, this required a pixel array of at least 1,067 px. Therefore, a resolution of 1080P, i.e. 1,920$\times$1,080 px, met our experiment requirements. In the same way, the frame rate was set to allow for particle tracking from one frame to another. We limited this requirement to the accurate tracking of the mm-sized beads during the shaking events, as tracking individual JSC-1 grains in these phases has proven to be difficult\citep{colwell2008Icarus} and was considered to be outside of the scope of this experiment. In order to track particles reliably throughout the experiment run, each particle should be seen moving less than its diameter from one frame to the other. To track the motion of a bead of 2~mm in diameter from one frame to another, while it is moving at a speed of 5 cm/s (expected maximum particle speed during shaking events), the time between two frames has to be 40~ms or less. Therefore, a frame rate of 30 fps was acceptable. The camera was set to record continuously. 
\enlargethispage{\baselineskip}

The shortest possible time between shaking events was set by the corresponding time unit (seconds, minutes, hours) that we hardcoded on the NCM. We chose the time unit to be minutes as we expected particle settling times of $>$200~s and thus did not deem it necessary for shaking events to be more frequent than a few minutes. This assumption was calculated from the evolution profile of the average collision particle velocity, $v=v_0\epsilon^N$, $v_0$ being the average initial velocity of the particles after each shaking event, $\epsilon$ the average coefficient of restitution, and $N$ the average number of collisions that happened to each particle. An estimation of the coefficient of restitution was obtained from a simple laboratory experiment, in which a bead was dropped ~50 times onto a surface covered with the same type of beads. High-speed cameras recorded the rebound of the dropped bead and the coefficient of restitution of the collision was determined for each test. For the acrylic beads and at the lowest drop heights (lowest collision velocities of $<$10 cm/s), the average measured coefficient of restitution was ~0.9. This was the highest compared to the glass and copper beads. Therefore, the calculated settling times calculated for the acrylic beads represented an upper limit to and we initially set the spacing between shaking events to 3 minutes. For an expected initial velocity of $v_0$$\sim$5 cm/s (induced by a shaking event) and an estimated coefficient of restitution of $\epsilon$=0.9, we calculated that average particle velocities would decrease to under 1~mm/s after 38 collisions and under 0.1~mm/s for 59 collisions. The time $T_N$ required for $N$ collisions to happen to a particle depends on the average collision frequency $f_n$ after the $n^{th}$ collision, where $f_n$ is given by $f_n=v_n/\lambda$, with $v_n=v_0\epsilon^n$ the average particle velocity after the $n^{th}$ collision and $\lambda$, the average particle mean free path. Therefore, 
\begin{equation}
T_N=\sum\limits_{n=0}^N\frac{1}{f_n}=\sum\limits_{n=0}^N\frac{\lambda}{v_0\epsilon^n}=\frac{\lambda}{v_0}\times\sum\limits_{n=0}^N(\frac{1}{\epsilon})^n=\frac{\lambda}{v_0}\times\frac{1-(\frac{1}{\epsilon})^{N+1}}{1-\frac{1}{\epsilon}}
\end{equation}
For the 15 2~mm in diameter acrylic beads in their 15$\times15\times3$~mm$^3$ experiment cell, we have $\lambda = (n\sigma)^{-1} = 1.5$~cm, $\sigma = \pi r^2$ being the cross section of each particle, with $r$ = 1 mm the radius of the acrylic beads. With our estimate of $v_0$ and $\epsilon$, we calculated the time needed for 38 and 59 collisions to happen to be $T_{38} = 162$~s and $T_{59} = 1,500$~s, respectively. 

In the continuous recording mode, the camera started recording upon power-on for a chosen overall duration, dividing the recorded data into a chosen number of video data files. Both the overall duration of the recording and the length of the data chunks were indicated in the NCM parameter file. We could change these parameters during the experiment's time on-board ISS by uploading a text file to the NCM. As described in the previous paragraph, during the development phase of the payload we calculated the expected particle settling time after each shaking event to range from 200 to 2,000~s. Therefore, we chose an overall recording duration of 60 minutes. The limitations imposed by ISS power and communication resources did not allow for a continuous (24/7) operation of the NanoRocks payload during its time on-board the space station. Instead, discrete periods of operations were planned every few months.

As an unexpected power-off of the camera during a recording would produce a corrupted video file and result in the loss of that entire data chunk, we set the individual video file length to 60~seconds. An unexpected loss of power to the payload would therefore only erase the last minute of recording. Data transfer from the NanoRocks experiment to UCF was by way of planned routine data downlink from ISS; retrieving the NanoRocks payload including its micro SD card was not required for the success of the experiment. 

\section{NanoRocks experiment runs onboard the International Space Station}
\label{s:run}

\subsection{Observed Particles}

We chose particle samples to maximize the scientific return of the experiment. They are listed in Figure~\ref{f:particles}: red spherical acrylic beads, blue aspherical glass beads, aspherical copper beads, and angular JSC-1 lunar simulant grains. The acrylic beads were 2~mm in average diameter, the glass and copper beads were 1~mm in average diameter, and the JSC-1 particles had a size distribution around 0.75 mm in diameter. Additionally, some particles were coated with a layer of fine chalk powder.

The size and composition of the particles were chosen to have particle masses around a few milligrams. The  acrylic, glass, and copper beads, and the JSC-1 grains at mm sizes have masses ranging from 1 to 35 mg. The density of the beads were 1.18, 2.6, 8.96, and 2.9 g/cm$^3$ for acrylic, glass, copper, and JSC-1, respectively. At the relative velocities induced by the shaking mechanism of the experiment (1 cm/s and below), observing this particle size allowed us to monitor the transition between bouncing and sticking collisions (see Figure~\ref{f:model}). In addition, we varied the shape and surface of the particles. The acrylic beads were spherical, the glass and copper beads were aspherical (ellipsoidal), the JSC-1 grains were angular, and we coated the particles in trays 2 and 4 with a chalk layer.

These different particle properties allowed us to determine the influence of (1) particle size (by comparing the behavior between beads and JSC-1 grains in trays 1 and 8, for example), (2) particle shape (by comparing the spherical beads of tray 1 with the non-spherical beads of trays 3 and 5, as well as the JSC-1 grains of tray 8), (3) coefficient of restitution (by comparing between trays 1, 3, and 5), and (4) surface texture (by comparing between coated and non-coated beads in trays 2 and 3, and 4 and 5) on the collision behavior. Due to the limited space available on this platform, the influence of some factors had to be combined. In particular, it will be difficult to distinguish the influence of size and shape between trays 1 and 8 on the particle behavior and additional experiments will be required in the future.

The observation of spherical and ellipsoidal particles is of particular interest to be able to compare our experimental results to numerical simulations reproducing the same initial conditions. Being able to calibrate such simulations with experimental data will allow for more accurate simulations of dust grain behavior at PPD size scales.

\subsection{Experiment Sequence}
\label{s:exp_sequence}

Each experiment sequence consisted of a series of experiment runs during which the samples were observed while they were shaken. At the start of an experiment run, the micro SD card was switched to the camera and the camera powered on. The recording then started automatically and lasted 60 minutes, dividing the data into 60-second video files. During an experiment run, the particles were agitated regularly.  

After the first data returned from ISS, we recognized that the damping time for the particle velocities after each shaking event was $\sim$20~s. We then set the shaking interval to its minimum, 60~s, and all further experiment runs were performed with this shaking interval. Once each 60-minute recording was completed, the NCM board cut the power to the camera and LED array until the start of the next experiment run. The microSD card was then switched back to the NCM to grant access from the ISS laptop. 

We first chose to run the experiment 3 times in a row every other day to complete one experiment run. This would allow us to check on the scientific data after a week and, if necessary, update the NCM parameter file for future experiment runs. A new experiment run was triggered by a power cycle of the payload, so that every power cycle reinitiated the NCM and restarted an experiment sequence.

\subsection{Experiment Performance}

\begin{figure}[t]
  \begin{center}
  \includegraphics[width = 0.5\textwidth]{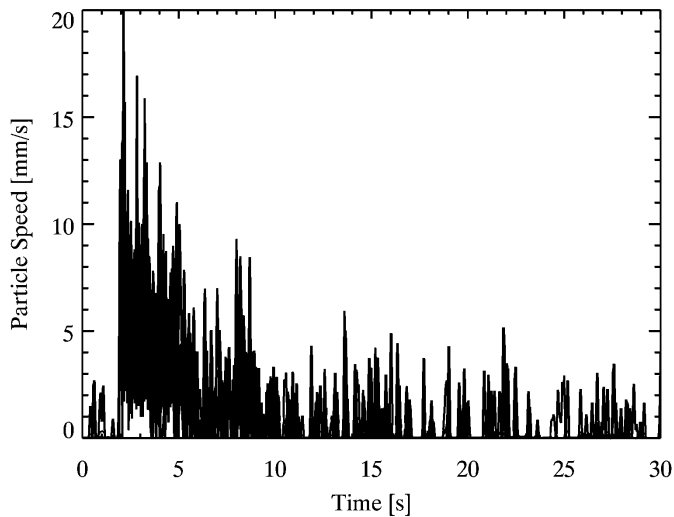}
 \caption{Instantaneous particle velocities determined by manual tracking for 10 particles of tray 1 (10 superposed curves). The shaking event occurs at t = 2 s in this graph.}
 \label{f:tray1_vel}
 \end{center}
\end{figure}

\begin{figure}[t]
  \begin{center}
  \includegraphics[width = 0.5\textwidth]{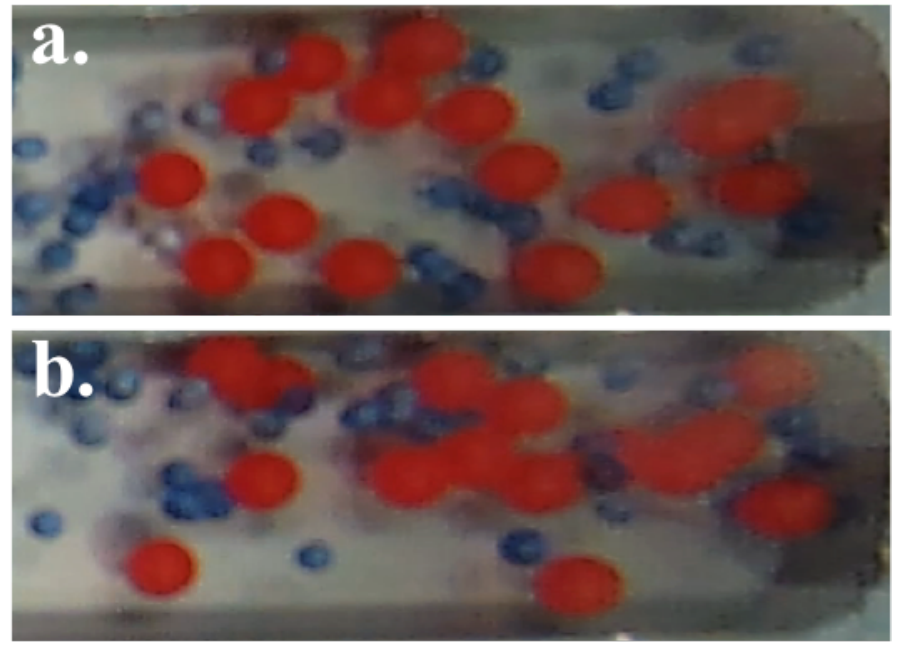}
 \caption{Particles in tray 3: (a) 0.1 s after the shaking event, and (b) 20 s after the shaking event. In (a) the particles are moving about as individual beads. What could appear as sticking contacts on this still frame are actually rebound collisions. In (b) particles have formed clusters of acrylic (red) and glass (blue) particles. Not all particles were involved in clusters and individual left-over particles remained as they became immobile (velocities $<$ 1 mm/s). The video recording of a full experiment run can be found at youtube.com/watch?v=x\_QmLBT25JA.}
 \label{f:clustering}
 \end{center}
\end{figure}

The video data we received from the payload demonstrated the nominal operation of all the experiment components, the LED arrays, camera, and shaking magnet. The particle velocities induced by the shaking events were measured by tracking a subset of particles using the manual tracking tool Spotlight \citep{spotlight}. Figure~\ref{f:tray1_vel} shows an example of the tracking results for 10 particles in tray~1 (acrylic beads, see Figure~\ref{f:particles}). The measured maximum particles velocities peaked at 2 cm/s with an average of $\sim$1 cm/s, which was somewhat lower than our goal of $\sim$5~cm/s. We were able to observe particle collisions at relative velocities decreasing from 1~cm/s to $<$1~mm/s (our motion detection threshold with the camera was around 1~mm/s). In addition, we observed the formation of particle clusters at very low collision velocities (under 1~mm/s). Figure~\ref{f:clustering} shows the particle clustering in tray 3 as an example. The only unexpected factor was the time the particle systems required to damp the velocities after each shaking event. We had predicted damping times of at least 200~s based on expected collision parameters, but the NanoRocks data showed that the particle velocities were significantly damped after only 20~s. This was due to a lower initial average particle velocity after each shaking event and average coefficients of restitution different than assumed.

\begin{figure*}[ht]
  \begin{center}
  \includegraphics[width = 0.95\textwidth]{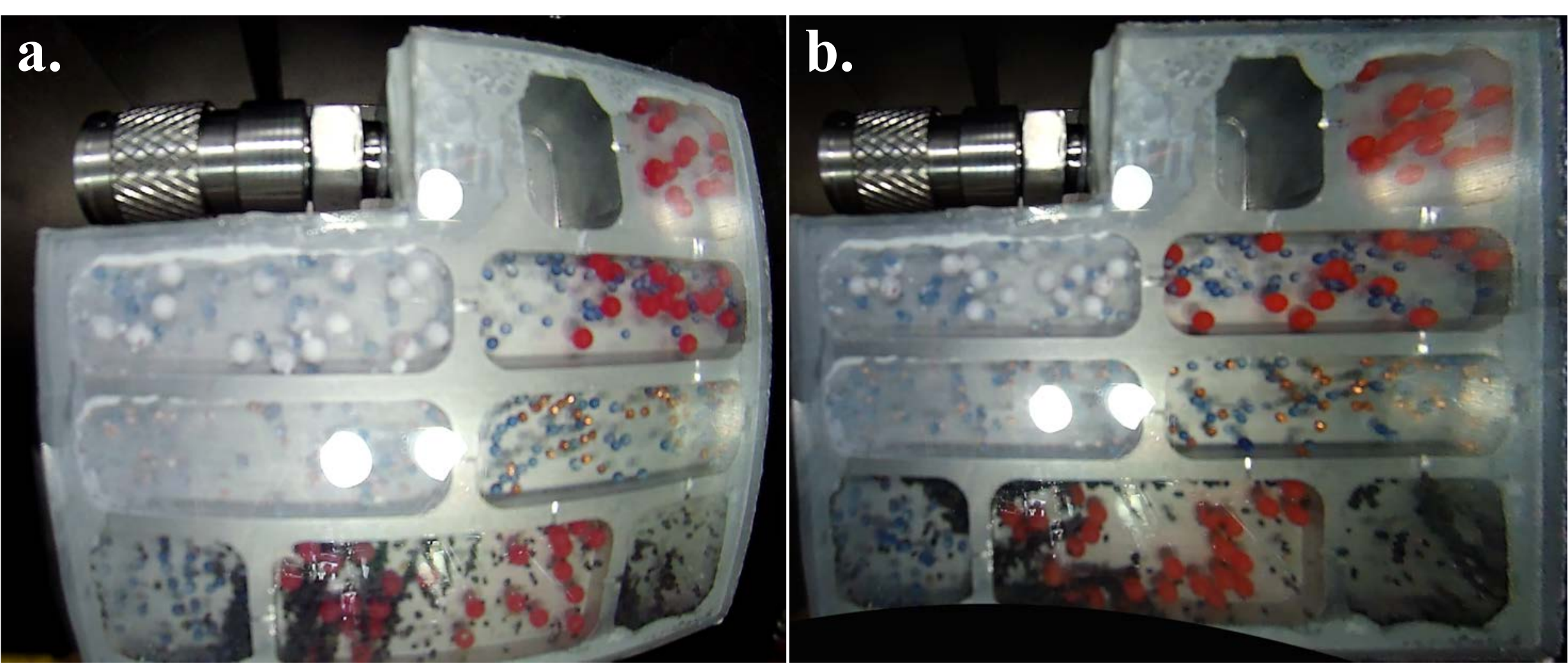}
 \caption{NanoRocks video data: (a) original frame distorted by the camera fisheye lens, (b) processed frame after distortion correction using GIMP.}
 \label{f:flight_data}
 \end{center}
\end{figure*}

As described above, the data collected by NanoRocks was in the form of video files. Figure~\ref{f:flight_data}a shows an example of a frame captured during an experiment run. In this image we see the distortion introduced by the fisheye lens of the HackHD camera. Figure~\ref{f:flight_data}b shows another frame after distortion correction using the GIMP software package. Three LEDs can also be seen reflected by the tray cover. The specular light reflection creates areas of the tray where particles cannot be seen thus reducing the area usable for data analysis. However, the surface affected is limited to three distinct and immobile spots, which allows for easy elimination during image processing. A light diffusing screen was considered but not implemented due to concerns that launch vibrations would dislodge it and obstruct the camera. Data analysis from this experiment will be published separately.

During its stay on-board the ISS we received a total of 158 video files from the NanoRocks payload through direct download from the ISS laptop. The experiment recorded more data, but not all the files could be downlinked to ground due to limitations in the available downlink time (the ISS downlink is a resource shared with other facilities). Out of the 158 video files received, the two first videos were from the first experiment sequence and each had a duration of 3 minutes. The other 156 videos were each 60~s long due to adjustment of the shaking interval and video length after the first sequence. The NanoRocks payload was retrieved from ISS and returned to the CMR, where we had the opportunity to retrieve the flight SD card from the UCF board to collect the remaining video files. Unfortunately, the SD card was damaged upon payload disassembly, therefore no further data could be retrieved from the flight hardware. 

The amount of data retrieved from NanoRocks was considerably higher than the amount of data collected from previous microgravity particle collision experiments \citep{brisset2016, weidling_et_al2012Icarus, kothe_et_al2013Icarus}. These former experiments were flown on platforms with limited available microgravity time (a few seconds in the Bremen drop tower, a few minutes on suborbital rockets). In fact, the amount of data produced by NanoRocks obliged our team at UCF to develop new and automated data analysis methods for the video analysis. Previous data analysis could be performed manually (particle tracking in particular) due to the limited amount of data collected, but manual analysis was not a viable option for the amount of data produced by NanoRocks. Automated tracking has enabled us to validate statistical analysis of the evolution of particle velocities \citep{colwell2015DPS,metzger2016ASCE}.

\section{Summary and Conclusion}
\label{s:conclusion}

The NanoRocks payload was a 1.5U experiment designed to study low-energy collisions in multi-particle environments that flew on the International Space Station from September 2014 to March 2016. In this paper, we described the science objectives, experiment setup, and performance of the payload in order to illustrate the possibilities that orbital miniaturized payloads offer for planetary science.

During its time on-board the ISS, NanoRocks functioned nominally and collected over 158 video files of scientific data. This very successful experiment also allowed us to benefit from several lessons learned. In addition to small hardware caveats like the lack of a light diffuser, we gained insight into the multi-particle collisional environment and thus ideas on how to optimize follow-up experiments. In particular, the upcoming CubeSat Particle Aggregation and Collision Experiment (Q-PACE) developed at the CMR will study multi-particle collisional systems on an orbital platform and will benefit from the hardware experience and data analysis performed with NanoRocks. Finally, it can be noted that the microgravity platform of the ISS was perfectly suited for the NanoRocks experiment and might be used in the future for further multi-particle collision experiments.


\begin{acknowledgements}
This work is based in part upon research supported by NASA through the Origins of Solar Systems Program grant NNX09AB85G and by NSF through grant 1413332. The NanoRocks experiment was supported by Space Florida, the ISSRDC, NanoRacks LLC, and the University of Central Florida.
\end{acknowledgements}

\bibliography{literatur}

\end{document}